\documentclass[twoside]{article}
\usepackage{qic,epsfig}

\usepackage{graphicx}
\usepackage{subfigure}
\usepackage{bbold}

\newcommand{\beq}{\begin{equation}}
\newcommand{\eeq}{\end{equation}}
\newcommand{\bea}{\begin{eqnarray}}
\newcommand{\eea}{\end{eqnarray}}
\newcommand{\bra}[1]{\left\langle #1 \right\vert}
\newcommand{\ket}[1]{\left\vert #1 \right\rangle}

\newcommand{\F}{{F}}

\newcommand{\unityop}{\hat{\mathbb{1}}}

\newcommand{\zerol}{\ket{0_{\rm L}}}
\newcommand{\onel}{\ket{1_{\rm L}}}
\newcommand{\brazerol}{\bra{0_{\rm L}}}
\newcommand{\braonel}{\bra{1_{\rm L}}}
\newcommand{\gen}{\ket{Q}}
\newcommand{\genortho}{\ket{Q_\perp}}
\newcommand{\bragen}{\bra{Q}}
\newcommand{\bragenortho}{\bra{Q_\perp}}
\newcommand{\rhogen}{\gen \bragen}
\newcommand{\rhogenortho}{\genortho\bragenortho}
\newcommand{\rhotrans}{\hat{\varrho}}
\newcommand{\rhotransortho}{\hat{\varrho}_\perp}
\newcommand{\alp}{\alpha}
\newcommand{\ang}{\phi}

\newcommand{\synd}[1]{\ket{S_{#1}}}
\newcommand{\brasynd}[1]{\bra{S_{#1}}}

\newcommand{\poo}{p_{00}}
\newcommand{\poI}{p_{01}}

\newcommand{\poT}{p_{02}}

\newcommand{\rhogenl}{\hat{\rho}(\incorrect)}
\newcommand{\rhogenorthol}{\hat{\rho}_\perp(\incorrect)}
\newcommand{\rhocel}[1]{\rho_{\rm c #1}}
\newcommand{\rhoc}{\hat{\rho}_{{\rm c}}}
\newcommand{\rhos}{\hat{\rho}_{\rm s}}
\newcommand{\rhor}{\hat{\rho}_{\rm r}}
\newcommand{\flip}{p_f}

\newcommand{\incorrect}{\gamma}
\newcommand{\digterm}{P}
\newcommand{\unitary}[1]{\hat{U}(#1)}
\newcommand{\codeword}{code word}
\newcommand{\codewords}{code words}
\newcommand{\codespace}{code space}
\newcommand{\signflip}{sign flip}
\newcommand{\bitflip}{bit flip}
\newcommand{\flipchannel}{flip channel}
\newcommand{\fliperror}{flip error}

\begin{document}
\setlength{\textheight}{8.0truein}    

\runninghead{Fidelity as a figure of merit in quantum error correction}
            {Jonas Alml\"{o}f and Gunnar Bj\"{o}rk}

\normalsize\textlineskip
\thispagestyle{empty}
\setcounter{page}{1}

\copyrightheading{0}{0}{2003}{000--000}

\vspace*{0.88truein}

\alphfootnote

\fpage{1}

\centerline{\bf
FIDELITY AS A FIGURE OF MERIT IN QUANTUM ERROR CORRECTION}
\vspace*{0.37truein}
\centerline{\footnotesize
JONAS ALML\"{O}F\footnote{jalml@kth.se}}
\vspace*{0.015truein}
\centerline{\footnotesize\it Department of Applied Physics, Royal Institute of Technology (KTH), AlbaNova University Center}
\baselineskip=10pt
\centerline{\footnotesize\it Stockholm, S-106 91, Sweden}
\vspace*{10pt}
\centerline{\footnotesize
GUNNAR BJ\"{O}RK}
\vspace*{0.015truein}
\centerline{\footnotesize\it Department of Applied Physics, Royal Institute of Technology (KTH), AlbaNova University Center}
\baselineskip=10pt
\centerline{\footnotesize\it Stockholm, S-106 91, Sweden}
\vspace*{0.225truein}
\publisher{(Oct 29 2011)}{(May 9, 2012)}

\vspace*{0.21truein}

\abstracts{
We discuss the fidelity as a figure of merit in quantum error correction schemes. We show that when identifiable but uncorrectable errors occur as a result of the action of the channel, a common strategy that improves the fidelity actually decreases the transmitted mutual information. The conclusion is that while the fidelity is simple to calculate and therefore often used, it is perhaps not always a recommendable figure of merit for quantum error correction. The reason is that while it roughly speaking encourages optimisation of the ``mean probability of success'', it gives no incentive for a protocol to indicate exactly where the errors lurk. For small error probabilities, the latter information is more important for the integrity of the information than optimising the mean probability of success.
}{}{}

\vspace*{10pt}

\vspace*{3pt}
\communicate{to be filled by the Editorial}

\vspace*{1pt}\textlineskip    

\section{Introduction}
Quantum computing, i.e., physical operations on qubits and entangled qubits, has attracted interest since the '70s, see e.g. \cite{feynman} and the references therein. Early on, it was realised that a major obstacle for its implementation is the undesired, but unavoidable, interaction with the environment, described as a quantum channel. Since this interaction typically is not unitary, the channel-qubit interaction can not be ``undone'' using a single unitary transformation. However, it was discovered in \cite{shor} that using a specific coding, a set of conditional, unitary transformations, each invoked by the result of a cleverly chosen measurement that locates and identifies the error, could correct one Pauli bit-flip error, phase error, or a combination of the two.  In the footsteps of this discovery, many new quantum error correction codes (QECCs) were developed, including those based on the stabiliser formalism \cite{Stab1,Stab2,Stab3}, Calderbank-Shor-Steane (CSS) codes \cite{CSS1,CSS2}, decoherence-free subspace codes \cite{DFS1,DFS2}, channel adapted codes such as amplitude damping channel codes \cite{Stab3,leung}. Such QECCs utilise redundant information for their function. These QECCs have often been compared using fidelity, i.e., essentially the probability that the state after being acted on by the channel and corrected by the so-called recovery operations, is identical to the initial state. Therefore, this figure of merit (FOM) is a measure of ``likeness'', and it has the characteristic property that identical states have unit fidelity, while orthogonal states have zero fidelity. We have also presented a code with fidelity as its FOM \cite{almlofbjork}. However, in this work we point out some problems associated with the use of fidelity as a FOM for QECCs.

First we would like to comment on ``similarity measures'', i.e., measures that quantify how well one quantum state resembles another, and how they are fundamentally different from information integrity measures, such as Shannon's ``rate of transmission'' \cite{shannon} -- nowadays referred to as mutual information \cite{hamming,cover}. E.g, assume that Alice has three binary, classical channels to use for communicating with Bob, one with fidelity $F=1$, one with $F=0.5$ and one with $F=0$. At a first glance, the $F=1$ channel appears to be the best choice, but in fact, the $F=0$ channel is equally good, because Bob needs only to bit-flip the data stream to get the correct data sequence. In contrast, the $F=0.5$ channel is utterly useless (if this fidelity is due to random noise), since it gives a vanishing value of the mutual information between Alice and Bob. Conversely, the maximum value of the mutual information for a given channel achieved by optimising the input alphabet, equals to the channel capacity.

The quantum counterpart of mutual information is the quantum mutual information \cite{qmutualinformation_1,qmutualinformation_2}, and analogously to the situation above, for transmission of qubits, it vanishes for $F=0.5$. It is also true that a QECC with zero overall fidelity could indeed be a very useful one!

Quantum mutual information and its classical counterpart, are relative measures that quantify how much information two parties can agree upon. Mutual information relates two sets of outcomes linked by a joint probability distribution, and quantum mutual information relates two subsystems described by a bipartite density operator. Also, a connection between quantum mutual information and perfect error correction exists, as shown in \cite{Ogawa} and references therein.

Another illustrative example is given by Shannon \cite{shannon}: Assume a transmission of $1000$ (uncoded) classical bits per second - each bit taking the values $0$ or $1$ with probability $1/2$, and also assume an error rate of $1\%$, where the errors result with equal probabilities in $0$ and $1$ regardless of the original bit value. A reader not familiar with information theory may be tempted to guess that the rate of transmission will decrease with $1\%$, becoming $990$ bits per second. However, the loss of transmission rate (equivocation) in this case is significantly larger, about $8.1\%$. This is due to the lack of knowledge of \emph{where} the errors are located. This knowledge suggests that in the case of detectable but uncorrectable errors, the faulty bits should be discarded and their locations ``tagged''. The error rate would then be minimised to about $10$ bits per second. Note that had we wanted to optimise ``similarity'' between the sent and received bit-strings, we could in the case of detectable errors ``improve'' the erroneous bits by replacing them by random bits with equal probabilities. This would cause the number of identical bits between sender and receiver to become $995$ per second on average. However, this optimisation of similarity will not increase the rate of information transmission at all. This illustrates the seemingly odd fact that the integrity of each bit is more important than how similar the input symbol strings are to the output ones.

Criteria for the construction of QECCs exist for the case where errors are reversible, i.e., expressed in terms of Pauli $X$- and $Z$ operators \cite{QECconditions,knill} where, given a maximum number of such errors $e$, perfect recovery of a coded logical qubit can be achieved. If, on the other hand, logical \codeword{}s are allowed to interact with a reservoir system, such errors can approximately be corrected if criteria given in \cite{leung} are satisfied, reaching a fidelity of the same order (in the error rate) as the reversible case. Other codes take advantage of the fact that in addition to correcting $e$ errors, any occurrence of $e+1$ errors can also be detected, so that such uncorrectable qubits can be discarded \cite{knill}.

When more than $e$ errors occur for a coded qubit, the ensuing state either fall into a space orthogonal to the \codespace{}, or overlap it. If such an erroneous state belongs to the orthogonal space, it can be identified but not corrected. There are then two conceivable strategies: One is to apply some recovery operator $\mathcal{R}_i$ that effectively replaces the erroneous state with a predefined state. The other strategy is to discard the state and take note of the location of the error (i.e., to ``tag'' it).

If the code is not perfect, i.e., if the \codespace{} does not completely occupy the extended space including all possible errors, one is left a possibility to recover additional errors. How these recovery operators should be constructed is not immediately given by the code criteria, but has to be established as the optimum recovery for a particular FOM, typically maximising fidelity for a fixed assumed statistical distribution of channel input states, and a given encoding and channel interaction \cite{optchannelfid,yamamoto,Fletcher1}. If, in addition, the error rate is known, the FOM can be improved significantly \cite{Stab3}. In these contexts the optimisation FOMs used are fidelity, entanglement fidelity, minimum fidelity or average fidelity. When discussing QECC in the context of quantum computers, other measures of closeness has been introduced \cite{knill}, particularly in connection to estimating errors from additional sources, such as quantum gate errors. It has been shown that for trace-preserving completely positive maps, optimising such closeness measures, e.g., trace distance, is akin to optimising fidelity \cite{optchannelfid, schumacher_approximate}. However, closeness measures such as fidelity can be deceitful as we have exemplified above.

Below we show that optimisation of the mutual information does not lead to optimisation of ``probability of sameness'', i.e., fidelity, and vice versa. In Sec. \ref{Sec: Classical} we shall compare the two figures of merit for a simple classical channel, using two different strategies of handling errors. We shall then repeat the calculation for a quantum channel in Sec. \ref{Sec: Quantum treatment}. As for the channel, we will assume that for some of the error events, all information about the error type was lost in the channel interaction. We compare two strategies that deal with such errors, one devised in \cite{Rahn} where the scheme applies the identity operator to map the error back to the \codespace{} -- in doing so optimising fidelity  -- and one where the qubit is discarded and its location is ``tagged'' and transmitted to the receiver. We shall show that the two strategies lead to opposite results, the former gives a higher fidelity than the latter, while the latter gives a higher mutual information than the former.

\section{Classical treatment}
\label{Sec: Classical}

To study the effects of the two strategies in a classical setting, we consider errors caused by \bitflip{}ping. A sequence of bits is sent through a channel which influences the individual bits independently, and causes them to flip, $0 \leftrightarrow 1$, with some probability $\flip$ per unit time.

Assume we have a classical binary channel and that at a certain time $t=0$ the probability that the bit is incorrect is $\incorrect (0)$. This means that between the times $t$ and $t+dt$, the change in probability of having the incorrect state will be
\beq
d \incorrect = - \incorrect \flip dt + (1-\incorrect) \flip dt .
\eeq
Going to the limit $dt \rightarrow 0$ and solving the ensuing differential equation one finds that
\beq
\incorrect(t) = \left (\incorrect(0)- \frac{1}{2} \right ) \exp(-2 \flip t) + \frac{1}{2} .
\eeq

Below we shall assume that after passing through the \flipchannel{},
each bit is in the incorrect state with probability $\incorrect$, where
$0 \leq \incorrect \leq 1/2$. We shall also assume that the bits flip
independently of each other, so that the probability of having,
e.g., four bits all in the correct state after passing through the
channel is $(1-\incorrect)^4$, and having exactly three out of four bits in
the correct state is $4 \incorrect (1-\incorrect)^3$.

To demonstrate our thesis, we will first consider
error correction in this classical channel model. Assume, e.g., that we code a
logical zero onto the four bit string $0000$, and the logical one
onto the string $1111$. This code can correct single \bitflip{} errors,
e.g., $0010$ will be interpreted as a logical zero, and the code can identify
two \bitflip{} errors, e.g., $0011$ can either be the result of
flipping the last two bits of $0000$ or the first two bits of $1111$. If
three errors occur, then the string will be incorrectly interpreted
as correctable and the ``correction'' will result in an erroneous bit. We write the
channel model for such a channel in Table \ref{Table: Symmetric}.
\begin{table}[h]
\tcaption{The channel matrix for a symmetric \flipchannel{}.}
    \centerline{\footnotesize\smalllineskip
    \begin{tabular}{|c|c|c|c|}
\hline
 Input $x$ & \multicolumn{3}{|c|}{Output $y$}\\
    \hline
 & 0 & 1 & Uncorrectable \\
\hline
0 & $\poo$ & $\poI$ & $\poT$ \\
\hline
1 & $\poI$ & $\poo$ & $\poT$ \\
 \hline
    \end{tabular}}
\label{Table: Symmetric}
\end{table}
For simplicity we have assumed that the channel is symmetric, that is, both the
\bitflip{}s and the coding is symmetric with respect of the
permutation $0 \leftrightarrow 1$. The probability of obtaining the
correct bit after decoding is denoted $\poo=(1-\incorrect)^4 + 4 \incorrect (1-\incorrect)^3 $, the probability for detecting an
uncorrectable string (e.g., $0101$) is denoted $\poT=6 \incorrect^2 (1-\incorrect)^2 $, and the
probability for erroneous correction is $\poI=\incorrect^4 + 4 \incorrect^3 (1-\incorrect) $. An asymmetric
channel will yield quantitatively different but qualitatively
similar results.

Using this channel model we shall investigate two strategies. The
first, called I, is to randomly assign the value zero and the value one when
we find that the \codeword{} we get cannot be corrected. Hence, the
effective new channel matrix is shown in Table \ref{Table: Concatenated}.
\begin{table}[h]
\tcaption{The effective channel matrix for a random assignment of bit-value when the error cannot be corrected.}
    \centerline{\footnotesize\smalllineskip
    \begin{tabular}{|c|c|c|}
\hline
 Input $x$ & \multicolumn{2}{|c|}{Output $y$}\\
    \hline
 & 0 & 1  \\
\hline
0 & $\poo+\frac{\poT}{2}$ & $\poI+\frac{\poT}{2}$  \\
\hline
1 & $\poI+\frac{\poT}{2}$ & $\poo+\frac{\poT}{2}$  \\
 \hline
    \end{tabular}}

\label{Table: Concatenated}
\end{table}
We see from the table that this strategy will increase the
probability of success from $\poo$ to $\poo + \poT/2$. At the same
time the probability of error will increase from $\poI$ to $\poI +
\poT/2$. If we were to optimise the probability for success, which
is akin to maximise the fidelity between the received and corrected
state in a quantum error correction scheme, then this would be a good strategy.

The transmitted mutual information $I$ is defined \beq I(X,Y) = S(X)
+ S(Y) - S(X,Y), \label{Eq: Mutual info} \eeq where $S(X)$ is the
entropy function \beq S(X) = - \sum_x p(x) \log_2[p(x)] \label{eq: entropy} \eeq of the random
variable $X$ expressed in bits. Using the properties of logarithmic
function and probability distributions, the mutual information can
be expressed in the simple form \beq I(X,Y) = -\sum_{x,y} p(x,y)
\log_2[\frac{p(x) p(y)}{p(x,y)}]. \label{Eq: Simpler expression
mutual info}\eeq We shall assume that we transmit bits, that is
$p(x=0)=p(x=1)=1/2$. The assumption of symmetric channel then leads
to $p(y=0)=p(y=1)=1/2$. From the table and the assumptions we also
get $p(0,0) = p(1,1)= \poo+\poT/2$ and $p(0,1) = p(1,0)=
\poI+\poT/2$. The mutual information in bits can now readily be
computed as \beq I_I = 1+\left ( \poo+\frac{\poT}{2}\right )\log_2
\left ( \poo+\frac{\poT}{2}\right )+\left (
\poI+\frac{\poT}{2}\right )\log_2 \left ( \poI+\frac{\poT}{2}\right
).\eeq

\begin{figure}[htbp]
\centerline{\epsfig{file=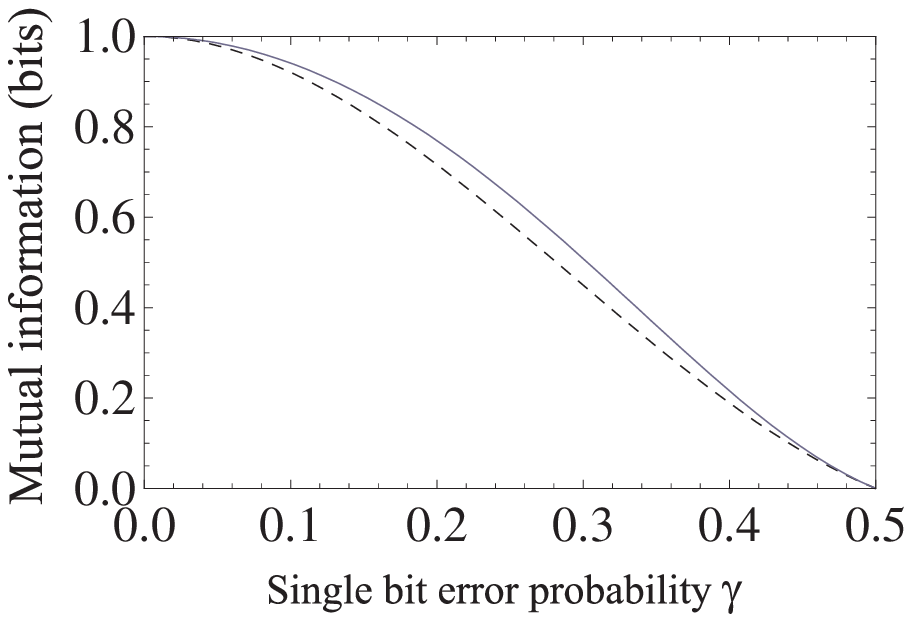, width=8.2cm}} 
\vspace*{13pt}
\fcaption{\label{fig:1a}The mutual information between a sent logical bit, and the coded, received, and corrected bit, as a function of the probability that a bit passing through the \flipchannel{} becomes incorrect. Strategy I is drawn dashed and strategy II drawn solid.}
\end{figure}

\begin{figure}[htbp]
\centerline{\epsfig{file=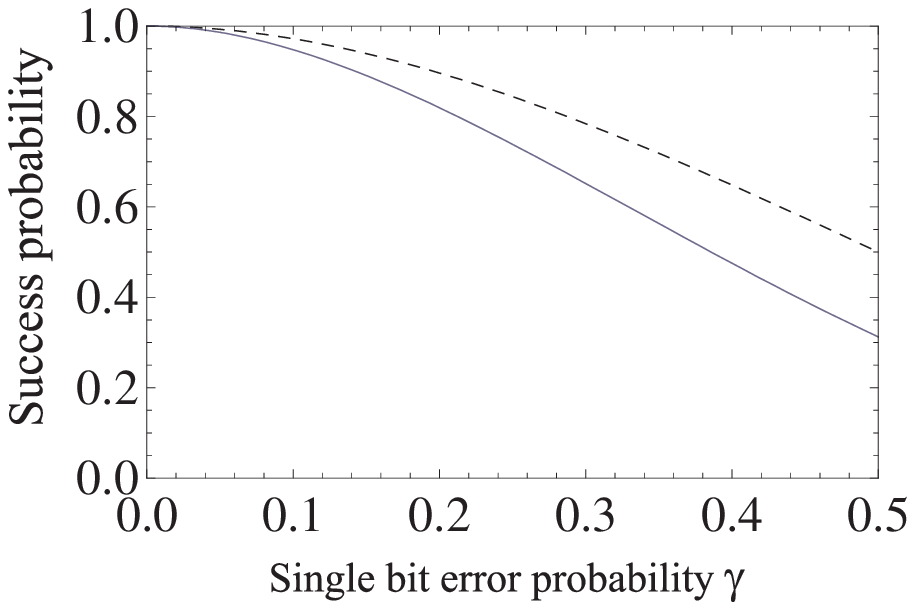, width=8.2cm}} 
\vspace*{13pt}
\fcaption{\label{fig:1b}The probability that a sent logical bit, and the coded, received, and corrected bit have the same values, as a function of the probability that a bit passing through the \flipchannel{} becomes incorrect. Strategy I is drawn dashed and strategy II drawn solid.}
\end{figure}

Another strategy, called II, is to simply discard any string that is
identified to be uncorrectable. If the information is to be
transmitted further, this information is irrevocably lost, but any such string
will be identified by appropriate tagging, e.g., by sending a
message on a separate channel pointing out which bit strings should
be ignored. The channel model \textit{for the ``successful'' bits that are left after this ``error
correction''} is displayed in Table \ref{Table: Method II}.
\begin{table}[h]
\tcaption{The effective channel matrix for the bits that are corrected, after bits that have been determined as erroneous but uncorrectable have been tagged as such and discarded.}
    \centerline{\footnotesize\smalllineskip
    \begin{tabular}{|c|c|c|}
\hline
 Input $x$ & \multicolumn{2}{|c|}{Output $y$}\\
    \hline
 & 0 & 1  \\
\hline
0 & $\frac{\poo}{1-\poT}$ & $\frac{\poI}{1-\poT}$  \\
\hline
1 & $\frac{\poI}{1-\poT}$ & $\frac{\poo}{1-\poT}$  \\
 \hline
    \end{tabular}}

\label{Table: Method II}
\end{table}
In this case, quite trivially, the overall probability for success is $\poo
\leq \poo + \poT/2$. The marginal distributions will remain the
same, but the mutual information of each of these ``successful'' bits is now \beq
I_{{\rm OK}} = 1+ \left ( \frac{\poo}{1-\poT}\right )\log_2 \left (
\frac{ \poo }{1-\poT}\right )+\left ( \frac{\poI}{1-\poT}\right
)\log_2 \left ( \frac{\poI}{1-\poT}\right ).\eeq However, in a long
string of bits, a fraction $\poT$ of the transmitted and coded bits are
going to be tagged to be discarded or ignored. Hence, the average
transmitted mutual information is \beq I_{\rm II} = (1-\poT) I_{{\rm OK}}. \eeq
A plot of $I_{\rm II}$ and  $I_{\rm I}$ is shown in Fig. \ref{fig:1a}. It
is seen that $I_{\rm II} \geq I_{\rm I}$, and that it is only equal for the
totally deterministic and the totally random channel. In Fig.
\ref{fig:1b} the success probability is plotted, that is, the
classical counterpart to the fidelity between the input and output
bits. For strategy II we assign zero success if the bit is either
incorrectly ``corrected'' or tagged as uncorrectable. It is clear that
for protocol II, where the probability of success is lower than for
protocol I, the transmitted mutual information is higher. This
indicates that optimisation of the success probability of a error
correction protocol does not lead to a (simultaneous) maximisation of the mutual
information, and vice versa, in general.

Next we shall show that the same conclusion holds for a quantum channel.

\section{Quantum treatment}
\label{Sec: Quantum treatment}

In the following, to exemplify our assertion, we shall
look at a specific channel and a specific code. For simplicity we shall look at a qubit \flipchannel{} which is a special case of a Pauli channel. The code we shall employ is the $[[1,5,3]]$ perfect code discovered independently in \cite{paz} and \cite{5qubit}. This code uses five physical qubits to code one logical qubit and it can correct one Pauli error, that is a \bitflip{}, a \signflip{}, or a combination of both. These errors are operationally described by the Pauli operators $\hat{\sigma}_x$, $\hat{\sigma}_z$, and $\hat{\sigma}_y$, hence the name.

One implementation of the $[[1,5,3]]$ code uses the following encoding \cite{paz}:
\bea \zerol & \rightarrow & \synd{000} = (-\ket{00000} + \ket{01111} - \ket{10011} + \ket{11100} \nonumber\\
& & + \ket{00110} + \ket{01001} + \ket{10101} + \ket{11010})/\sqrt{8} \nonumber \\
\onel & \rightarrow & \synd{100} = ( -\ket{11111} + \ket{10000} + \ket{01100} - \ket{00011} \nonumber \\
& & + \ket{11001} + \ket{10110} - \ket{01010} -
\ket{00101})/\sqrt{8} \label{eq: code}\eea
The \codewords{} are
orthogonal and the code is designed such that that \bitflip{}ping
($\ket{0}\leftrightarrow\ket{1}$) any one of the five qubits will
give an additional $2 \times 5$ state vectors that are mutually
orthogonal, and also orthogonal to the \codewords{}. Likewise,
\signflip{}ping ($-\ket{1}\leftrightarrow\ket{1}$) or simultaneously
flipping both the bit-value and the sign
$\pm\ket{0}\rightarrow\pm\ket{1}$ and
$\pm\ket{1}\rightarrow\mp\ket{0}$) of any of the five qubits
will result in twenty vectors that are mutually orthogonal, and also
orthogonal to the $12$ previous vectors. Hence, these $32$ vectors,
called syndrome vectors, will form an orthonormal basis in the
$2^5=32$-dimensional Hilbert \codespace{}. We will use the notation
$\synd{jkl}$, where the index $j$ denotes the logical qubit ($0$ or
$1$), $1 \leq k \leq 5$ indicates which qubit is erroneous, and $l=1,2,3$ denotes that a $\hat{\sigma}_x$-,
$\hat{\sigma}_z$-, or $\hat{\sigma}_y$-flip has occurred, respectively. The corresponding eigenvalue is denoted $S_{jkl}$.

In the following we will, for the sake of reasoning, assume the
following model: A long string of states, randomly selected between
two orthogonal qubit states $\gen$ and $\genortho$ is generated in
duplicate. (Should the states not be selected with equal probability
our conclusions below would still hold qualitatively, but would be
quantitatively somewhat different.) In the ensemble sense, the
generated states are thus described by the following density matrix: \beq \hat{\rho} = \frac{1}{2} \left ( \gen \otimes \gen
\bragen  \otimes \bragen + \genortho \otimes \genortho \bragenortho \otimes
\bragenortho \right ) . \label{eq: initial state}\eeq One
of the duplicate states is subsequently encoded onto a five-qubit
state according to the encoding (\ref{eq: code}). The coded state is sent
to a receiver through a \flipchannel{} where any qubit will be flipped
with probability $\incorrect$ independently of any other qubit. On the
receiver side, the five-qubit states are measured by a quantum non-demolition, von Neumann
measurement having the \codewords{} and the syndrome vectors as
eigenstates, with pairwise degenerate eigenvalues. The states
$\synd{000}$ and $\synd{100}$ form one such pair, and the syndrome vectors
$\synd{0kl}$ and $\synd{1kl}$ constitute the $15$ other pairs.
We shall then apply one of two different strategies with the measured five-qubit state:
\begin{romanlist}
    \item[I] Decode it by a measurement-result specific unitary operation, unless the measurement collapsed the vector onto one of the vectors $\synd{jk2}$ and $\synd{jk3}$, $j=0,1$, $k=1, \ldots , 5$, in which case the erroneous state is randomly replaced by either $\zerol$ or $\onel$. In the ensemble sense it is replaced with the maximally mixed state $(\zerol\brazerol + \onel\braonel)/2$.
    \item[II] Decode it by a measurement-result specific unitary operation, unless the measurement collapsed the vector onto one of the vectors $\synd{jk2}$ or $\synd{jk3}$, $j=0,1$, $k=1, \ldots , 5$, in which case the qubit is flagged as uncorrectable and is discarded.
\end{romanlist}

The motivation behind the two strategies is as follows: In strategy
I we use the $[[1,5,3]]$ code, but since the code is designed for
the Pauli channel, which is more general than the \flipchannel{} we
have assumed, it is not optimal. We know from the code's
construction, that a single \bitflip{} in a \codeword{} will always
result in a state that is orthogonal to the states $\synd{jk2}$ and
$\synd{jk3}$, $j=0,1$, $k=1, \ldots , 5$. Hence, if the measured
vector collapses onto one of the syndrome vectors $\synd{jk2}$ or $\synd{jk3}$ we know that
there has been more than one \fliperror{}. This implies that we cannot
correct the vector, as the code is designed to correct only single
errors. In order to increase the fidelity between the final vector
and the sent vector we shall randomly replace the measured vector
with $\zerol\brazerol$ or $\onel\braonel$. In an ensemble sense,
the measured state is replaced by the density
matrix $(\zerol\brazerol + \onel\braonel)/2$. A motivation for this
strategy is, e.g., given in \cite{Rahn}: "As the initial state ...
is known to be in the \codespace{}, it is clearly more beneficial to
return the state ... to the \codespace{} than do otherwise: lacking any
other information one could at least prepare the completely mixed
state in the \codespace $(\ket{0}\bra{0} +
\ket{1}\bra{1})/2$, yielding an average fidelity of $1/2$,
rather than leaving the register outside the \codespace{}, yielding the
fidelity of 0." We have used the same strategy to increase a code's fidelity in \cite{almlofbjork}.

Strategy II does exactly that \cite{Rahn} argues
against, namely when the five-qubit state is projected onto an uncorrectable syndrome vector, it is identified as uncorrectable and sent outside
the \codespace{}. In this case, since the code and syndrome vectors
span the whole space, only the null vector remains. The result is
exactly the one Rahn {\em et al.} predicts, the average fidelity
becomes lower than using strategy I, but as we shall
see below, {\em the average quantum mutual information between the sent and
error corrected qubit is higher than for
strategy I.}

A general logical-qubit state can be written \beq \gen = \sin \alp
\zerol + e^{i \ang} \cos \alp \onel . \label{Eq: general state}\eeq
The state orthogonal to $\gen$ is \beq \genortho = \cos
\alp \zerol -  e^{i \ang}  \sin \alp \onel . \label{Eq: general orthogonal state}\eeq
Using (\ref{eq: code}), (\ref{eq: initial state}), (\ref{Eq: general
state}), (\ref{Eq: general orthogonal state}), and the channel model
we can compute the joint density matrix after the coded state has
been subjected to the \flipchannel{} and been corrected using either
strategy I or II. The probability of having no flip will be
$(1-\incorrect)^5$, and the probability of having qubit $1$, $2$ and $3$ flipped
is $\incorrect^3(1-\incorrect)^2$. In the second case, the \codeword{} $\zerol$
will become \beq (-\ket{11100} + \ket{10011} - \ket{01111} +
\ket{00000} + \ket{11010} + \ket{10101} + \ket{01001} +
\ket{00110})/\sqrt{8} .\eeq However, it is straightforward to compute that this is identically the same
state as the syndrome $\synd{113}$, that is the syndrome for the
case when the first qubit of \codeword $\onel$ has undergone a
simultaneous bit and \signflip{}. Hence, the three-bit flipped state
$\zerol$ will be detected as erroneous but uncorrectable. Depending on strategy, this state will subsequently be
``corrected'' to become $(\zerol\brazerol + \onel\braonel)/2$
according to strategy I, or detected as erroneous and uncorrectable, and be discarded
according to strategy II.

Assume that the state $\gen \bragen = \hat{\rho}(0)$ was sent. After being transmitted through the channel it becomes $\rhogenl$. Suppose this state is measured by a quantum non-demolition detector in the syndrome vector basis, and that the syndrome measurement resulted in the (degenerate) eigenvalue associated with the indices $kl$. The state then collapses into the unnormalised density matrix
\beq
\rhotrans_{kl} = \left(
  \begin{array}{cc}
    \brasynd{0kl}\rhogenl\synd{0kl} & \brasynd{0kl}\rhogenl\synd{1kl} \\
    \brasynd{1kl}\rhogenl\synd{0kl} & \brasynd{1kl}\rhogenl\synd{1kl} \\
  \end{array}
\right)
\label{eq:measuredstate}
\eeq
when expressed in the $\left\{ \zerol,\onel \right\}$ basis. The trace of $\rhotrans_{kl}$ is equal to the probability of obtaining the measurement results $S_{0kl}$ or $S_{1kl}$.
If the syndrome has $l=1$ the state is correctable via the unitary transformation $\unitary{k} = \zerol\brasynd{0k1} + \onel\brasynd{1k1}$. If the syndrome has $l=2,3$ it is not correctable, and one can show that the measured syndrome gives no clue as how to correct the state. Summing up the undisturbed, and the correctable contributions, one arrives at the density matrix
\beq
\rhotrans = \rhotrans_{00} + \sum_{k=1}^5 \rhotrans_{k1}.
\label{eq:correctable_part}
\eeq
The probability to receive an uncorrectable state is $\digterm(\rhogenl)=\sum_{k=1}^{5} \sum_{l=2}^{3}(\brasynd{0kl}  \rhogenl \synd{0kl} + \brasynd{1kl}  \rhogenl \synd{1kl})$. Replacing $\rhogenl$ in (\ref{eq:measuredstate}) and (\ref{eq:correctable_part}) by $\rhogenorthol$ gives the state matrix $\rhotransortho$ and the probability $\digterm(\rhogenorthol)$ of not being able to correct the state, given that $\rhogenortho=\rhotransortho(0)$ was sent. Then the symmetry of the code implies that $\digterm(\rhogenl)=\digterm(\rhogenorthol)$ in this case.
The overall matrix ensuing from strategy I, given that the sent matrix is $(\rhogen \otimes \rhogen + \rhogenortho \otimes \rhogenortho)/2$ is thus a block diagonal matrix, that we can symbolically express
\beq
\rhoc=
\frac{1}{2}\left(
  \begin{array}{cc}
    \hat{\rho} & 0 \\
    0 & \hat{\rho}_{\perp} \\
  \end{array}
\right)\label{eq:joint_state}
\eeq
in the basis $\left\{ \zerol \otimes \gen, \onel \otimes \gen, \zerol \otimes \genortho, \onel \otimes \genortho \right\}$, where the diagonal elements can be identified as
\beq
\hat{\rho} = \rhotrans + \digterm(\rhogenl{}) \unityop /2, \qquad \hat{\rho}_{\perp} = \rhotransortho + \digterm(\rhogenl{}) \unityop/2.\label{eq:diagonal_uncorrectable_part}
\eeq
The identity operator in (\ref{eq:diagonal_uncorrectable_part}) represents the projection onto the $(\zerol\brazerol + \onel\braonel)/2$ state.

Strategy I now dictates that we have ${\rm Tr}(\rhoc)=1$ due to the completeness of the code- and syndrome vectors. From the assumption of equal probability in the transmitted sequence of vectors $\gen$ and $\genortho$ we have that $\rhocel{11}+\rhocel{22}=\rhocel{33}+\rhocel{44}=1/2$. Due to the symmetry of the code under the action of \bitflip{}ping we also have $\rhocel{11}+\rhocel{33}=\rhocel{22}+\rhocel{44}=1/2$. Hence, the qubits $\zerol$ and $\onel$ are received with equal probability.

In the case of strategy II, {\em for all states that are not flagged as uncorrectable}, one instead gets $\digterm(\hat{\rho})=0$ and using this $\digterm$, one gets ${\cal N}=\rhocel{11}+\rhocel{22}+\rhocel{33}+\rhocel{44}$. The fraction of these correctable states in a large ensemble will be ${\cal N}$. Again, due to the symmetry of sent state, channel, and code, the mutual information between the sender's and the receiver's error corrected and decoded qubits is given by Eq. (\ref{eq: simplified mutinfo}).

\begin{figure}[htbp]
\centering
\subfigure[]
{
\includegraphics[width=0.3\textwidth]{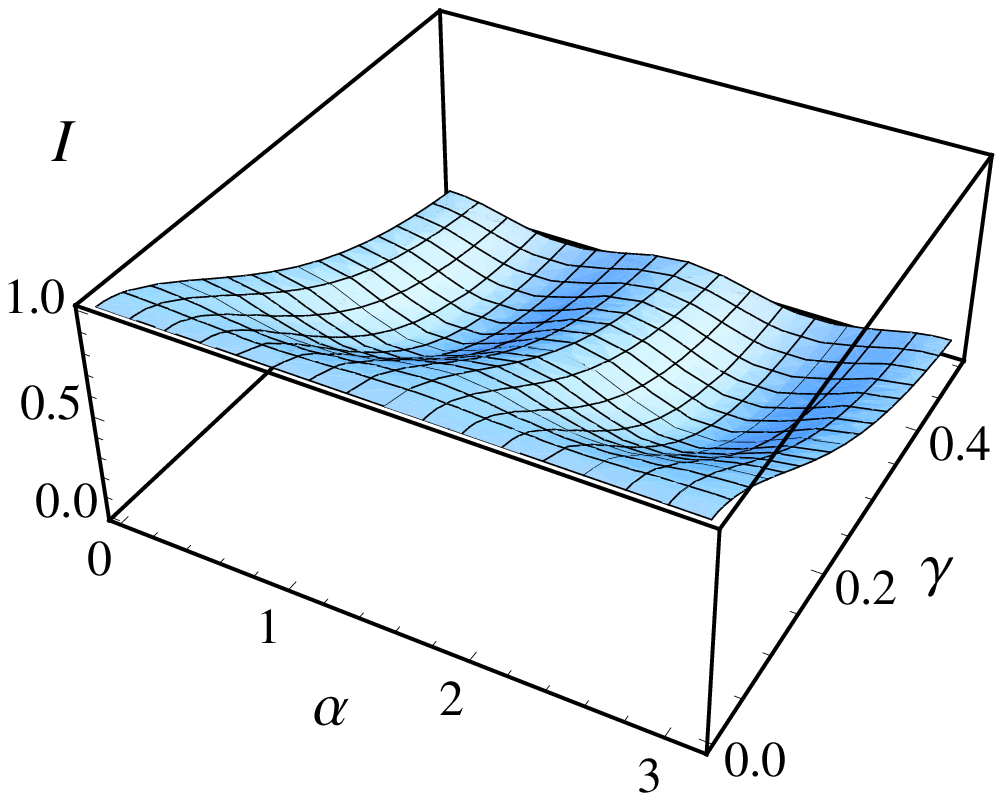}
    \label{fig:2a}
}
\hfill
\subfigure[]
{
\includegraphics[width=0.3\textwidth]{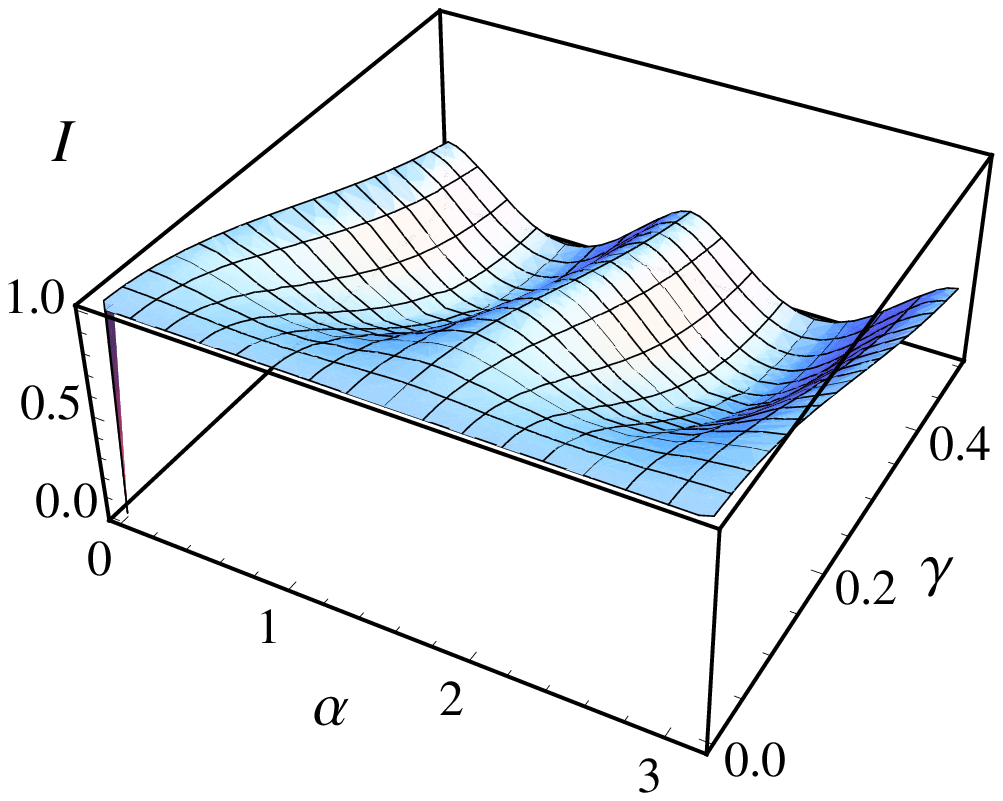}
    \label{fig:2c}
}
\hfill
\subfigure[]
{
\includegraphics[width=0.3\textwidth]{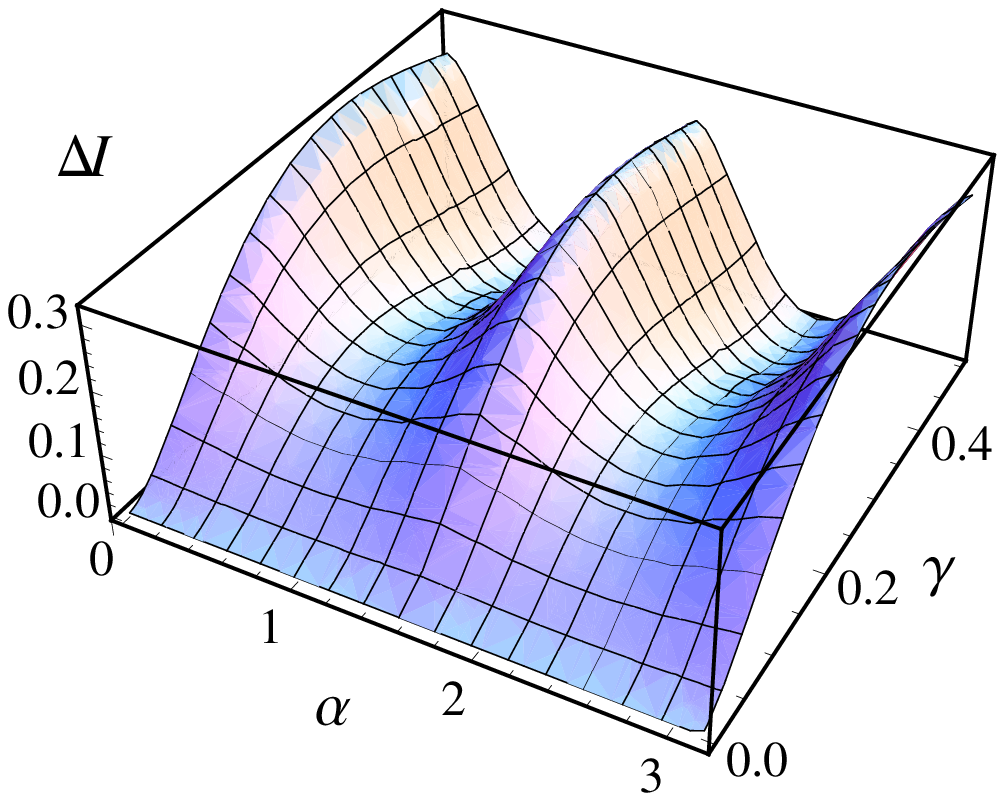}
    \label{fig:2e}
}
\hfill
\subfigure[]
{
\includegraphics[width=0.3\textwidth]{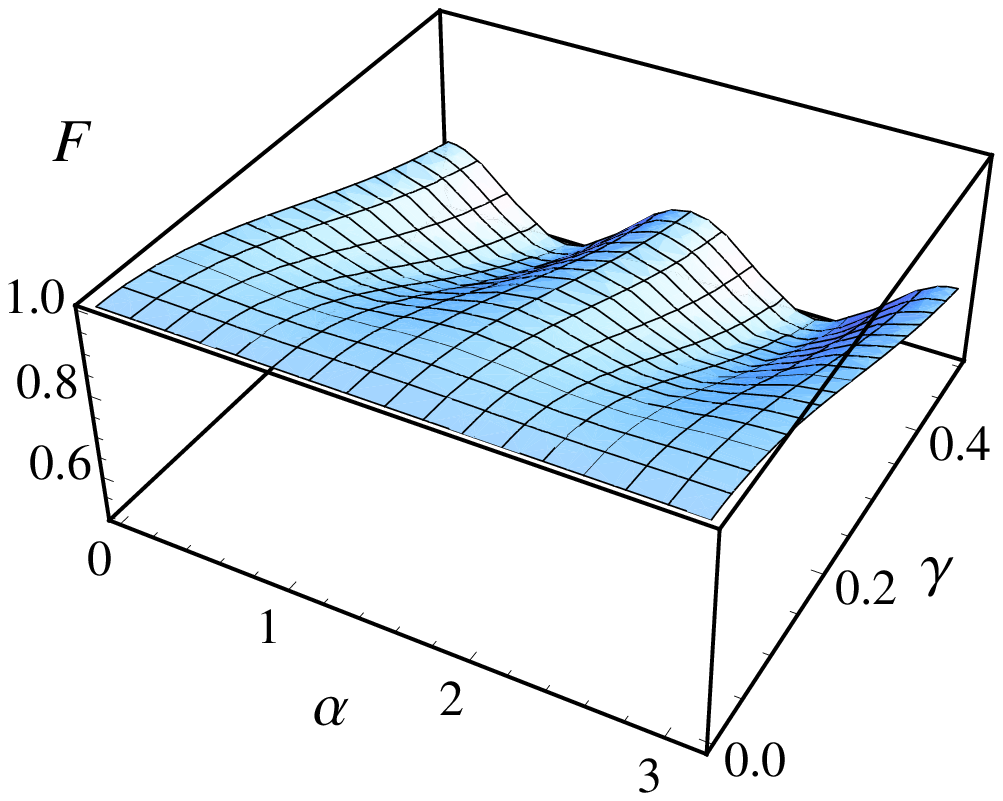}
    \label{fig:2b}
}
\subfigure[]
{
\includegraphics[width=0.3\textwidth]{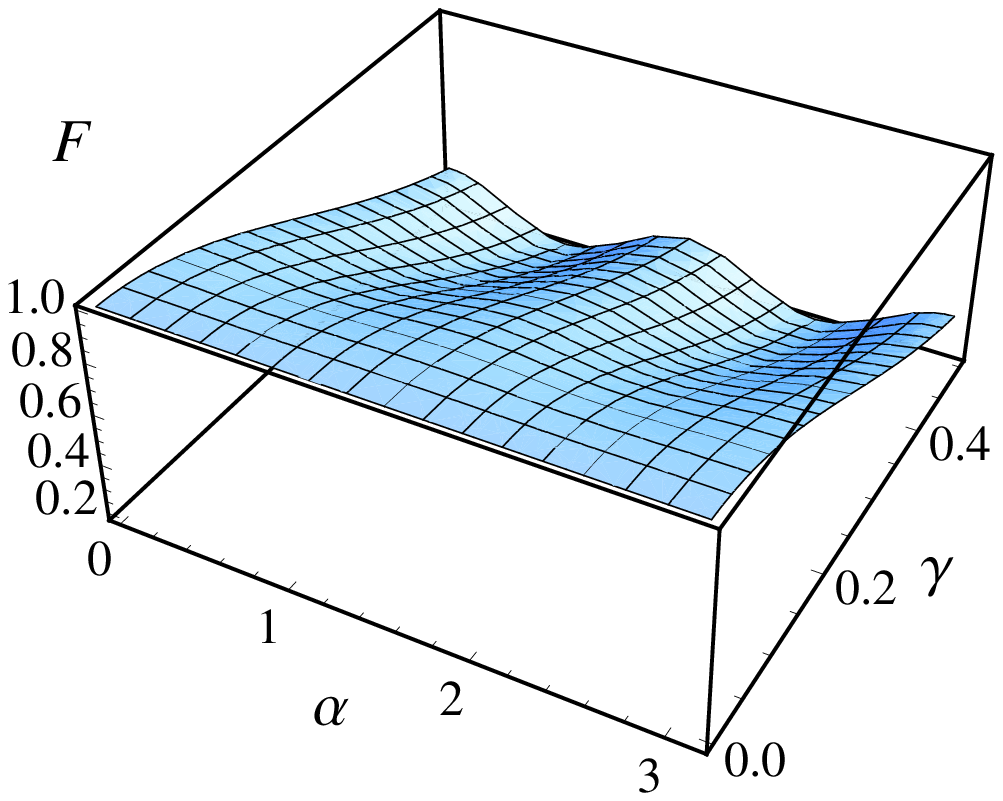}
    \label{fig:2d}
}
\subfigure[]
{
\includegraphics[width=0.3\textwidth]{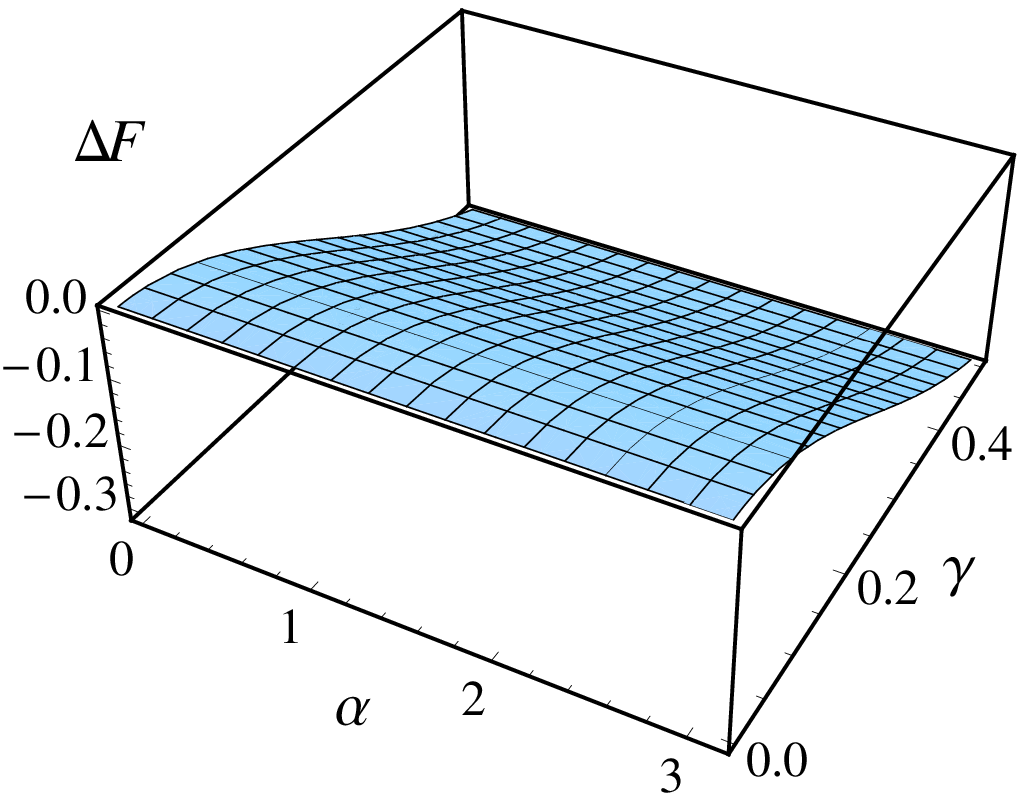}
    \label{fig:2f}
}
\fcaption{The mutual information between the sender's and the receiver's error corrected and decoded qubits (in bits), using strategy I in (a), strategy II in (b) and the difference, $I(\alpha,\gamma)_{\rm II}-I(\alpha,\gamma)_{\rm I}$ in (c). The average fidelity between the sender's qubit and the received and decoded qubit, using strategy I in (d), strategy II in (e) and the difference $\F(\alpha,\gamma)_{\rm II}-\F(\alpha,\gamma)_{\rm I}$ in (f).}
\label{fig:comparison}
\end{figure}

The mutual information between the sender's and the receiver's error corrected and decoded qubits (in bits) is \beq I(\incorrect,\alpha,\phi) = S(\rhos)+S(\rhor) - S(\rhoc{}) = 1 + S(\rhor) - S(\rhoc{}) , \label{eq: simplified mutinfo}\eeq where \beq S(\hat{\rho}) = - \rm{Tr}[\hat{\rho} \log_2(\hat{\rho})] \eeq is the von Neumann entropy (in bits) and $\rhos$ and $\rhor$ are obtained by tracing out the receiver and sender system, respectively, from $\rhoc{}$. The unity in the simplified, right-hand expression of (\ref{eq: simplified mutinfo}) comes from the fact that the sent qubits have the average entropy $1$ bit. A numerical evaluation of this expression yields Fig. \ref{fig:2a}, which shows the mutual information as a function of $\alp$ and $\incorrect$. (The function is quite naturally symmetric with respect to the line $\incorrect = 1/2$). The function is very weakly dependent on $\ang$ and the figure is drawn for $\ang=0$.

The average fidelity between the sender's qubit and the received and decoded qubit is \beq \F = {\rm Tr}\left ( \sqrt{\sqrt{\rhos} \rhor \sqrt{\rhos}} \right ) , \eeq
where
\beq \rhos = \frac{1}{2}\left(
\begin{array}{cc}
1 & 0 \\
0 & 1 \\
\end{array}
\right)
\eeq  and \beq \rhor=\left(
\begin{array}{cc}
\rhocel{11}+\rhocel{33} & \rhocel{12}+\rhocel{34} \\
\rhocel{21}+\rhocel{43} & \rhocel{22}+\rhocel{44} \\
\end{array}
\right)
\eeq
in the $\left\{ \gen , \genortho  \right\}$ basis.
Due to the simple form of $\rhos$, one gets $\F = (\sqrt{\lambda_1}+\sqrt{\lambda_2})/\sqrt{2}$, where $\lambda_1$ and $\lambda_2$ are the eigenvalues of $\rhor$. A numerical evaluation of this expression yields Fig. \ref{fig:2b}.

 However, the received information, averaged over all sent states, is given by ${\cal N} I(\incorrect)$ since the uncorrectable states carry no information. A numerical evaluation of the expression ${\cal N} I(\incorrect)$, valid for strategy II, yields Fig. \ref{fig:2c}.

The average fidelity in this case is obtained in the same manner as for strategy I, except for the fact that the so obtained fidelity must be multiplied with the factor ${\cal N}$ to take into account the instances when the states are uncorrectable and hence do not contribute to the fidelity. A numerical evaluation of the average fidelity, valid for strategy II, yields Fig. \ref{fig:2d}.

\section{Conclusions}
We have shown for both a classical and a quantum channel that if one wants to optimise the fidelity between a sent string of (qu)bits and the string after it has been error correction coded, transmitted through a noisy channel, and its error correction syndrome has been measured, then (qu)bits that can be identified as erroneous but that cannot be corrected should be mapped back onto the \codespace{} using the identity operation. We called this strategy I. If instead, one would like to optimise the (quantum) mutual information between the sender and receiver in an error correcting context, the best strategy is to discard (qu)bits that can be identified as erroneous but that cannot be corrected. This strategy was called II. Strategy I results in a lower (quantum) mutual information than strategy II, while strategy II results in a lower fidelity than strategy I. This illustrates an important insight, namely that fidelity and mutual information are not necessarily positively correlated in quantum error correction schemes. Depending on what quantum information protocol one intends to implement, one may want to optimise one of these figure of merits, but this should be done knowing that, in general, it will be done at the expense of the other. Hence, while fidelity is quite straightforward to calculate in comparison to quantum mutual information, one should avoid drawing the conclusion that all modifications that result in a higher fidelity will also increase the mutual information (and vice versa). Our simple example shows that the two figures of merit in general require different strategies to optimise.

\nonumsection{Acknowledgements}
\noindent
This work was supported by the Swedish Research Council (VR) through its Linn\ae us Center of Excellence ADOPT.

\nonumsection{References}

\end{document}